\documentclass[pra,showpacs,nofootinbib]{revtex4}
\usepackage{hyperref}
\usepackage{amsmath}
\usepackage{amssymb}
\usepackage{amsthm}
\usepackage{graphics}
\usepackage{graphicx}
\usepackage{color}
\def\bra#1{\langle#1|}
\def\dyad#1#2{|#1\rangle\langle#2|}

\def\ket#1{|#1\rangle }
\def\proj#1#2{|#1\rangle_#2\langle#1|}

\def\Tr#1{\textrm{Tr}(#1)}

\def\HC{{\cal H}}

\def\MC{{\cal M}}

\def\SC{{\cal S}}

\def\UC{{\cal U}}

\def\endproof{{\hspace{\stretch{1}}$\blacksquare$}}

\newtheorem{thm1}{Theorem}
\newtheorem{thm0}[thm1]{Theorem}
\newtheorem{cor2}[thm1]{Corollary}
\newtheorem{lem1}[thm1]{Lemma}
\newtheorem{lem2}[thm1]{Lemma}
\newtheorem{lem3}[thm1]{Lemma}

\newtheorem{thm4}[thm1]{Theorem}

\newtheorem{thm8}[thm1]{Theorem}
\newtheorem{thm9}[thm1]{Theorem}

\begin{document}
\title{All unitaries having operator Schmidt rank 2 are controlled unitaries}
\author{Scott M. Cohen}
\email{cohensm52@gmail.com}
\affiliation{Department of Physics, Portland State University,
Portland, Oregon 97201}
\author{Li Yu}
\affiliation{Centre for Quantum Technologies, National University of Singapore, 3 Science Drive 2, Singapore 117543}

\begin{abstract}
We prove that every unitary acting on any multipartite system and having operator Schmidt rank equal to $2$ can be diagonalized by local unitaries. This then implies that every such multipartite unitary is locally equivalent to a controlled unitary with every party but one controlling a set of unitaries on the last party. We also prove that any bipartite unitary of Schmidt rank $2$ is locally equivalent to a controlled unitary where either party can be chosen as the control, and at least one party can control with two terms, which implies that each such unitary can be implemented using local operations and classical communication (LOCC) and a maximally entangled state on two qubits.  These results hold regardless of the dimensions of the systems on which the unitary acts.

\end{abstract}

\date{\today}

\maketitle
\section{Introduction}\label{sec1}

Unitary gates are essential for quantum information processing, hence it is important to find simple ways to implement them.  In this paper we consider unitaries acting on $P$ systems, where $P\ge 2$.\footnote{We assume that the dimension of the Hilbert space of each system remains unchanged after the action of the unitary.}  Generally, methods for implementing a given unitary are discussed in two different scenarios: \emph{local} and \emph{nonlocal}.  In the \emph{local} scenario, the given unitary acting on $P$ systems is decomposed as a product of simpler unitaries, each of which may act on any number of systems.  In the \emph{nonlocal} scenario, many different parties at remote locations could share some ancillary quantum state that may be entangled, and they do local operations and classical communication (LOCC) to implement the desired unitary.  This paper is developed for the nonlocal scenario, but it applies to gate decomposition in the local scenario as well, see the remarks in the Conclusions.

A common approach is to express unitaries using a sum of product operators:
\begin{equation}\label{eqn1a}
\UC = \sum_j A_j \otimes B_j \otimes C_j \otimes \cdots \otimes R_j,
\end{equation}
where $A_j$, $B_j$, etc. are local operators on the respective parties. A simple expansion would help us find ways to implement $\UC$, but to judge what is a simple expansion there are at least two different criteria. The first is to use a numeric measure: for example, the Schmidt rank \cite{Tyson}, defined as the smallest possible number of product operators that can be summed to obtain $\UC$.  The second is to see whether the local operators in the expansion are simple. A set of mutually orthogonal projectors would qualify under this criterion. If the expansion on one party involves only such projectors, then $\UC$ is a controlled unitary.  Another simple type of local operators is discussed in \cite{NLU}.  In this paper we find a connection between the two criteria: all Schmidt rank-2 multipartite unitaries are equivalent to controlled unitaries under local unitaries. The result of the bipartite case is slightly stronger, see Theorem~\ref{thm8} below. Our result can be viewed as a structure theorem for the class of Schmidt rank-$2$ nonlocal unitaries. It characterizes what these unitaries are in terms of product-operator expansions.  Since we have solved the simplest nontrivial case of Schmidt rank $2$, the next goal would be to characterize nonlocal unitaries of higher Schmidt rank. The forms of expansions in \cite{NLU} may help toward this goal.

Amongst the various types of nonlocal unitaries, controlled unitaries are in some ways most easily understood. In the bipartite case, they are one of the few classes of unitaries for which their capacity to create entanglement between the parts is relatively well understood \cite{WangSandersBerry}, and significant progress has been made toward understanding their entanglement cost, that is, the amount of entanglement that is required to implement them without bringing the various parts together in a single laboratory \cite{Soeda,StahlkeU}. (The result of \cite{StahlkeU} also applies to some other classes of unitaries not yet completely characterized.) Controlled unitaries also play an important role in quantum information theory, being one constituent in commonly used gate sets that are universal \cite{Barenco} for quantum computation \cite{Nielsen}, as well as being instrumental in the creation of graph states and cluster states \cite{Sch}, which find wide use in quantum communication protocols \cite{BB84} and quantum computation.

Perhaps the simplest controlled unitary is the controlled-not on two qubits $A$ and $B$,
\begin{align}\label{eqn1}
\UC_{cn\!ot}=\proj{0}{A}\otimes I^{(B)}+\proj{1}{A}\otimes \sigma_x^{(B)},
\end{align}
where $I^{(B)}$ is the identity operator and $\sigma_x^{(B)}=\ket{0}_B\bra{1}+\ket{1}_B\bra{0}$ is the usual Pauli operator, both acting on system $B$. If system $A$ starts out in the $\ket{0}_A$ state, the state of the full $AB$ system is unchanged, whereas if the initial state of $A$ is $\ket{1}_A$, that system is unchanged, but the $B$ system is ``flipped". This notion is generalized to provide a definition of a controlled unitary, one for which if the input state of one system is a state in a given orthogonal basis, say the standard basis $\{\ket{j}_A\}$ as in the previous example, then that state is unchanged and unitary $W_j$ is performed on the remaining system, which may itself be multipartite,
\begin{align}\label{eqn2}
\UC=\sum_{j=1}^{d_A}\proj{j}{A}\otimes W_j.
\end{align}
Since it is easy for the controlling party to perform local unitaries on her system both before and after the action of $\UC$, we want to allow for this possibility. Then, every unitary $\UC$ for which there exist local unitaries $U^{(A)},V^{(A)}$ such that
\begin{align}\label{eqn3}
(U^{(A)}\otimes I^{(B)})\UC(V^{(A)\dag}\otimes I^{(B)})=\sum_{j=1}^{d_A}\proj{j}{A}\otimes W_j,
\end{align}
will be referred to as ``locally equivalent" to a controlled unitary with party $A$ controlling. This definition can be generalized in an obvious way to the case where multiple parties are controlling when $\UC$ acts on three or more parties, but we will require that the controlling parties each control locally, such as in
\begin{align}\label{eqn4}
\UC=\sum_{j,k}\proj{j}{A}\otimes\proj{k}{B}\otimes W_{jk}.
\end{align}
We will refer to a unitary acting on $P$ parties as ``fully controlled" if $P-1$ of the parties can act as controls. Our main result is

\begin{thm0}\label{thm0}
Every nonlocal unitary having Schmidt rank equal to $2$ is locally equivalent to i) a fully controlled unitary, and to ii) a diagonal unitary.
\end{thm0}
\noindent Note that i) and ii) are generally inequivalent without the Schmidt rank-$2$ assumption. The statement ii) implies i) with complete generality according to Lemma~\ref{lem1} below, but not the other way around.  For the special case of two qubits, however, i) and ii) are each equivalent to the unitary having Schmidt rank $2$, and therefore are equivalent to each other. The result i) for two qubits was previously obtained in \cite{Sch2IsControlledUTwoQubits}.

In the next section, we provide a series of results that are then used in section \ref{sec2b} to prove Theorem \ref{thm0}. Then, in section \ref{sec3} and for a unitary of any Schmidt rank operating on any number of parties, we provide a sufficient condition for when that unitary is locally equivalent to a controlled unitary, a condition that also tells us which party or parties can act as a control.

\section{Main Results}\label{sec2}
In this section we prove the result that every nonlocal unitary of Schmidt rank equal to $2$ is locally equivalent to a controlled unitary. We start out by proving a series of results that are then used to prove our main theorem.

\subsection{Preliminaries}\label{sec2a}
We begin with
\begin{lem1}\label{lem1}
A nonlocal unitary $\UC=\sum_jA_j\otimes B_j$, here written in a Schmidt expansion across the $A|B$ cut with $B$ itself possibly a multipartite system, is locally equivalent to a controlled unitary with party $A$ controlling iff the set of operators $\{A_j\}$ has a simultaneous singular value decomposition.
\end{lem1}
\proof By saying that a set of operators $\{A_j\}$ has a ``simultaneous singular value decomposition", we mean every operator in that set can be diagonalized by the same pair of unitaries, $U^{(A)}$ and $V^{(A)}$. In other words, $U^{(A)}A_jV^{(A)\dag}$ is diagonal for every $j$. To prove the ``if" part of this lemma, write $U^{(A)}A_jV^{(A)\dag}=\sum_{k=1}^{d_A}a_{jk}\proj{k}{A}$, from which we have that
\begin{align}\label{eqn10}
(U^{(A)}\otimes I^{(B)})\UC(V^{(A)\dag}\otimes I^{(B)})=\sum_jU^{(A)}A_jV^{(A)\dag}\otimes B_j=\sum_j\sum_{k=1}^{d_A}a_{jk}\proj{k}{A}\otimes B_j=\sum_{k=1}^{d_A}\proj{k}{A}\otimes\left(\sum_ja_{jk}B_j\right),
\end{align}
which indeed shows that this is a controlled unitary with $A$ controlling. 

To prove the converse, if $\UC$ is locally equivalent to a controlled unitary with $A$ controlling, then $\exists{U^{(A)},V^{(A)}}$ unitaries such that
\begin{align}\label{eqn11}
(U^{(A)}\otimes I^{(B)})\UC(V^{(A)\dag}\otimes I^{(B)})=\sum_{k=1}^{d_A}\proj{k}{A}\otimes W_k.
\end{align}
If $\UC=\sum_jA_j\otimes B_j$ is a Schmidt expansion, then the $\{B_j\}$ satisfy $\Tr{B_l^\dag B_j}=\delta_{jl}$, which implies from
\begin{align}\label{eqn12}
\sum_jU^{(A)}A_jV^{(A)\dag}\otimes B_j=\sum_{k=1}^{d_A}\proj{k}{A}\otimes W_k,
\end{align}
 that for each $l$,
\begin{align}\label{eqn13}
U^{(A)}A_lV^{(A)\dag}=\sum_{k=1}^{d_A}b_{lk}\proj{k}{A},
\end{align}
where $b_{lk}=\Tr{B_l^\dag W_k}$, and this completes the proof.\hspace{\stretch{1}}$\blacksquare$

The next lemma will also be useful.
\begin{lem2}\label{lem2}
Given $r$ maps $\{R_i\}$ from input space $\HC_{in}$ to output space $\HC_{out}$ such that the set of $r^2$ operators $\{R_i^\dag R_j\}$ spans a two-dimensional space containing the identity operator $I$, then there exist unitaries $U,V$ such that $UR_iV^\dag$ is diagonal for every $i$.
\end{lem2}
\noindent The proof is given in the appendix.

We will also use a result from \cite{mySumOfProd} that tells us when a sum of product operators can be equal to a product operator. This will be of use here, since we will be considering expansions of nonlocal unitary $\UC$ in terms of product operators, so that the unitary condition, $I=\UC^\dag\UC$, will appear as a sum of product operators that is equal to the product operator, $I$.
\begin{thm4}\label{thm4}
\cite{mySumOfProd} Given a set of product operators acting on two parties, $\{\MC_k=M_k^{(1)}\otimes M_k^{(2)}\}_{k=1}^N$, if there exists a set of nonzero coefficients, $\{c_k\}$, such that the linear combination $\SC=\sum_{k=1}^Nc_k\MC_k$ has Schmidt rank $r_s=1$ and so is also a product operator, then
\begin{align}\label{eqn92}
\delta_1+\delta_2\le N+1,
\end{align}
where $\delta_\alpha$ is the dimension of the space spanned by operators $\{M_k^{(\alpha)}\}_{k=1}^N$.
\end{thm4}
\noindent The connection between this theorem and Lemma \ref{lem2} is seen by considering $k$ as the composite index $(i,j)$, and $M_k^{(\alpha)}=R_i^{(\alpha)\dag}R_j^{(\alpha)}$. Since our aim is to analyze nonlocal unitaries, we will consider the case where $\SC$ in this theorem is the identity operator so that with $\UC=\sum_iR_i^{(1)}\otimes\ldots\otimes R_i^{(P)}$, $\{\MC_k\}$ provides a product expansion of $\UC^\dag\UC=I$.

We need one more lemma in order to prove our main theorem. This result is well known, but if the reader is interested in seeing a short proof, it can be found in \cite{mySumOfProd}.
\begin{lem3}\label{lem3}
If operators $\{M_j^{(\alpha)}\}_{j=1}^N$ span a space of dimension $\delta_\alpha$, then with $M_j^{(\beta)}\ne0~\forall{j}$ ($\beta\ne\alpha$), operators $\{M_j^{(\alpha)}\otimes M_j^{(\beta)}\}_{j=1}^N$ span a space of dimension no less than $\delta_\alpha$.
\end{lem3}

\subsection{Proof of Theorem \ref{thm0}}\label{sec2b}
The proof of Theorem \ref{thm0} follows directly from the results of the previous section. A unitary of Schmidt rank-$2$ acting on $P$ parties can be written
\begin{align}\label{eqn101}
\UC=M_1^{(1)}\otimes M_1^{(2)}\otimes \ldots\otimes M_1^{(P)}+M_2^{(1)}\otimes M_2^{(2)}\otimes \ldots\otimes M_2^{(P)},
\end{align}
with $M_j^{(\alpha)}$ an operator acting on party $\alpha$, and $\UC$ will satisfy
\begin{align}\label{eqn102}
I=\UC^\dag\UC=\sum_{i,j=1}^2 M_i^{(1)^\dag}M_j^{(1)}\otimes M_i^{(2)^\dag}M_j^{(2)}\otimes \ldots\otimes M_i^{(P)^\dag} M_j^{(P)}.
\end{align}
Let us first discuss the case that no terms on the right-hand-side of \eqref{eqn102} vanish. Then this expression is a sum of four product operators equal to a product operator, so that by Theorem \ref{thm4}, it must be that for any bipartite split $A|B$ of the $P$ parties, the spans of the corresponding operators on the two sides satisfy $\delta_A+\delta_B\le5$. This implies that no more than one party's set of operators $\{M_i^{(\alpha)^\dag}M_j^{(\alpha)}\}$, can span a space of dimension exceeding two. This can be seen as follows: suppose parties $\alpha$ and $\beta$ each have spans of dimension at least three. Consider a bipartite split where party $\alpha$ is in part $A$ and $\beta$ is in part $B$. By Lemma \ref{lem3}, $\delta_A\ge3$ and $\delta_B\ge3$, contradicting the above-stated requirement from Theorem \ref{thm4} that the sum of these cannot exceed $5$. Hence, for all parties but one, the spans have dimension no greater than two. If for a given party $\alpha$, $\delta_\alpha=1$, we must have $M_1^{(\alpha)}\propto M_2^{(\alpha)}$, and both these operators must be proportional to the same unitary. In this case, party $\alpha$ can just perform that unitary, which is uncorrelated to the actions on the other parties, and party $\alpha$ need not be considered further in the analysis. Then, we can just start over by considering a ``reduced" $\UC$ that only operates on the remaining parties. Alternatively, one may view party $\alpha$ as controlling, with the unitary that acts on the remaining parties being independent of the input on $\alpha$. If, on the other hand, $\delta_\alpha=2$, then by Lemma \ref{lem2}, there exist unitaries $U^{(\alpha)},V^{(\alpha)}$ acting on party $\alpha$ such that $U^{(\alpha)} M_1^{(\alpha)}V^{(\alpha)\dag}$ and $U^{(\alpha)} M_2^{(\alpha)}V^{(\alpha)\dag}$ are both diagonal, and this is true for every $\alpha$ for which $\delta_\alpha=2$. This is seen by partially tracing out all parties but one in \eqref{eqn102}, revealing that the spans of operators $\{M_i^{(\alpha)^\dag}M_j^{(\alpha)}\}$ contain the identity operator, and then the conditions of lemma \ref{lem2} are met for each party having $\delta_\alpha=2$. Hence for all parties but one, there exist local unitaries to diagonalize the $M_j^{(\alpha)}$ operators on that party. For each such party, there exists a set of orthogonal projectors such that the two diagonalized $M_j^{(\alpha)}$ are linear combinations of these projectors, see the proof of Lemma \ref{lem1}. Hence, all parties but one can act as controls simultaneously; that is, $\UC$ is ``fully controlled". In the expansion of $\UC$ using these projectors on $P-1$ parties, the operators on the remaining party are unitaries, and since $\UC$ is of Schmidt rank 2, these unitaries span a space of dimension $2$. Denote by $V_1$ and $V_2$ two of these unitaries that form a basis of this space. These become $I$ and $V_1^\dag V_2$ when $\UC$ is multiplied by $V_1^\dag$ on that party.  Then, $V_1^\dag V_2$ can be diagonalized under a unitary similarity transform, which does not alter the identity $I$, hence $I$ and $V_1^\dag V_2$ can be diagonalized simultaneously. Therefore, all the local operators (unitaries) in the expansion of $\UC$ on that last party can be simultaneously diagonalized, since they are all linear combinations of $V_1$ and $V_2$. Therefore $\UC$ is locally equivalent to a diagonal unitary.  This completes the proof of our main theorem for the case that there are no vanishing terms in \eqref{eqn102}.

The remaining case is that some terms in \eqref{eqn102} vanish, which only happens if for one party, say the first, $M_1^{(1)\dag} M_2^{(1)}=0=M_2^{(1)\dag} M_1^{(1)}$. If the corresponding operator products vanished for more than one party, then with Lemma \ref{lem3}, \eqref{eqn102} could not satisfy Theorem \ref{thm4}. This is because for a bipartite split $A|B$ with $A$ including only party $1$ and $B$ being all the rest, we have $\delta_A=2$ (because $M_1^{(1)\dag} M_1^{(1)}$ and $M_2^{(1)\dag} M_2^{(1)}$ cannot be proportional to each other while summing to the identity when $M_1^{(1)\dag} M_2^{(1)}=0$) implying from Theorem \ref{thm4} that $\delta_B\le N+1-\delta_A=1$, as $N=2$. Therefore, when $M_1^{(1)\dag} M_2^{(1)}=0=M_2^{(1)\dag} M_1^{(1)}$, $M_j^{(\alpha)}$ is proportional to a unitary for $j=1,2$ and $\alpha\ne1$, and
\begin{align}\label{eqn103}
\left(I^{(1)}\otimes M_1^{(2)\dag}\otimes M_1^{(3)\dag}\ldots\otimes M_1^{(P)\dag}\right)\UC=cM_1^{(1)}\otimes I^{(2)}\otimes\ldots\otimes I^{(P)}+M_2^{(1)}\otimes M_1^{(2)\dag} M_2^{(2)}\otimes\ldots\otimes M_1^{(P)\dag} M_2^{(P)},
\end{align}
for some constant $c$. Since $M_1^{(\alpha)\dag} M_2^{(\alpha)}$ is unitary for every $\alpha\ne1$, we can find unitaries $U^{(\alpha)}$ such that $U^{(\alpha)} M_1^{(\alpha)\dag} M_2^{(\alpha)}U^{(\alpha)\dag}$ is diagonal $\forall{\alpha\ne1}$. Then, by Lemma \ref{lem1}, all other parties can control the first, and the argument at the end of the last paragraph for $\UC$ being locally equivalent to a diagonal unitary still applies, completing the proof of Theorem \ref{thm0}.

\subsection{The bipartite case}\label{sec2c}
We will now show that
\begin{thm8}\label{thm8}
Any bipartite unitary of Schmidt rank $2$ is locally equivalent to a controlled unitary where either party can be chosen as the control, and at least one party can control with two terms.
\end{thm8}
\noindent For example, if the first party controls with two terms, then up to local unitaries
\begin{align}\label{eqn230}
\UC=P_1\otimes W_1+P_2\otimes W_2,
\end{align}
where $P_1$ and $P_2$ are orthogonal projectors.
\proof Given Theorem \ref{thm0}, it only remains to prove the claims that either party can control and that one of them can control with two terms. Assuming Theorem \ref{thm0} shows that it is the first party that can control, then 
\begin{align}\label{eqn16}
(U^{(1)}\otimes I^{(2)})\UC(V^{(1)\dag}\otimes I^{(2)})=\sum_{k=1}^{d_1}\dyad{k}{k}\otimes W_k.
\end{align}
Since $\UC$ has Schmidt rank $2$, it must be that the span of unitaries $\{W_k\}_{k=1}^{d_1}$ has dimension exactly equal to $2$. Choose ordering of the standard basis states on the first party's space such that $W_1,W_2$ are linearly independent. Then,
\begin{align}\label{eqn17}
W_k=\mu_{k1}W_1+\mu_{k2}W_2.
\end{align}
Inserting this into \eqref{eqn16}, we find
\begin{align}\label{eqn18}
(U^{(1)}\otimes I^{(2)})\UC(V^{(1)\dag}\otimes I^{(2)})=\sum_{k=1}^{d_1}\proj{k}{1}\otimes \sum_{j=1}^2\mu_{kj}W_j=\sum_{j=1}^2\left(\sum_{k=1}^{d_1}\mu_{kj}\proj{k}{1}\right)\otimes W_j:=\sum_{j=1}^2A_j\otimes W_j.
\end{align}
Since $W_k$ is unitary so that $W_k^\dag W_k=I^{(2)}$, then from \eqref{eqn17} we find that for every $k$,
\begin{align}\label{eqn19}
0=\mu_{k1}\mu_{k2}^\ast I^{(2)}-(1-|\mu_{k1}|^2-|\mu_{k2}|^2)W_1^\dag W_2+\mu_{k1}^\ast \mu_{k2}(W_1^\dag W_2)^2.
\end{align}
First let us assume that the coefficients in this quadratic equation for unitary $W_1^\dag W_2$ do not all vanish. Then when diagonalized, it becomes a quadratic equation for the eigenvalues of this unitary. Since all eigenvalues satisfy the same quadratic equation, which has exactly two distinct solutions, this means that $W_1^\dag W_2$ has exactly two distinct eigenvalues (which is why the quadratic equation cannot have only one distinct solution, since $W_1^\dag W_2\not\propto I^{(2)}$), and then with $P_1,P_2$ orthogonal projectors onto the degenerate subspaces of $W_1^\dag W_2$, thus providing a decomposition of the identity on the second party's space, $I^{(2)}=P_1+P_2$, we have
\begin{align}\label{eqn20}
W_1^\dag W_2=\lambda_1P_1+\lambda_2P_2,
\end{align}
and
\begin{align}\label{eqn21}
(U^{(1)}\otimes W_1^{\dag})\UC(V^{(1)\dag}\otimes I^{(2)})&=A_1\otimes (P_1+P_2)+A_2\otimes (\lambda_1P_1+\lambda_2P_2)\notag\\
&=Q_1\otimes P_1+Q_2\otimes P_2,
\end{align}
with $Q_j=(A_1+\lambda_jA_2)$ unitaries. Therefore, either party can control, and the second party can control with two terms.

We still need to consider the case that all coefficients in the quadratic \eqref{eqn19} vanish. Then for each $k$, either $\mu_{k1}=0$ and $|\mu_{k2}|=1$ or $\mu_{k2}=0$ and $|\mu_{k1}|=1$. In either case, we can multiply \eqref{eqn18} by the diagonal unitary $D=\sum_k\mu_{km}^\ast\proj{k}{1}$, with $m=1\textrm{ or }2$ chosen so that each diagonal element is nonzero, to obtain
\begin{align}\label{eqn22}
(DU^{(1)}\otimes I^{(2)})\UC(V^{(1)\dag}\otimes I^{(2)})=\sum_{j=1}^2\left(\sum_{k=1}^{d_1}|\mu_{kj}|^2\proj{k}{1}\right)\otimes W_j=P_1\otimes W_1+P_2\otimes W_2.
\end{align}
In this case, the first party can control with two terms. Multiplying this expression on the left by $I^{(1)}\otimes W_1^\dag$ and then performing a unitary similarity transformation on the second party to diagonalize $W_1^\dag W_2$, we see by Lemma \ref{lem1} that the second party can also control, completing the proof.\endproof

Theorem~\ref{thm8} has the following corollary:
\begin{cor2}\label{corr2}
Any Schmidt rank-$2$ bipartite unitary can be implemented using a maximally entangled state on two qubits and LOCC.
\end{cor2}
\noindent We omit the proof because this theorem follows directly from the results of \cite{NLU}, which also provides a simple protocol. Let us just offer a remark on whether this entanglement cost of 1 ebit is optimal: while it is shown in \cite{StahlkeU} that it is impossible to deterministically implement any Schmidt rank-$2$ bipartite unitary using a Schmidt rank-2 partially entangled state and LOCC, it may be possible by using a partially entangled state of Schmidt rank greater than $2$ with less than $1$ ebit of entanglement, see the example in \cite{StahlkeU}. Hence the optimal entanglement cost is not always $1$ ebit.

\section{Discussion: Unitaries of higher Schmidt rank}\label{sec3}
It is not difficult to find unitaries that are not controlled, and one need only go to Schmidt rank-$3$ to find simple examples. One such example, which acts on three parties, is
\begin{align}\label{eqn500}
\UC=\frac{1}{\sqrt{3}}\left(I\otimes I\otimes I+iX\otimes X\otimes X+iZ\otimes Z\otimes Z\right),
\end{align}
where $X$ and $Z$ are Hermitian unitaries that anticommute with one another, $XZ+ZX=0$. For such operators, no unitaries $U,V$ exist such that $UV^\dag,UXV^\dag,UZV^\dag$ are all diagonal (because this would require that $V^\dag U$ commutes with both $X$ and $Z$, and that $UXV^\dag$ commutes with $UZV^\dag$; together these commutation relations imply that $X$ commutes with $Z$). A common example of operators satisfying these conditions is the usual Pauli operators, $\sigma_x,\sigma_z$. Another well-known example, this time on a bipartite system and having Schmidt rank of $4$, is the SWAP operator on two qubits,
\begin{align}\label{eqn501}
\UC=\frac{1}{2}\left(I\otimes I+\sigma_x\otimes \sigma_x+\sigma_y\otimes \sigma_y+\sigma_z\otimes \sigma_z\right).
\end{align}

Of course, it is certainly also possible for unitaries of higher Schmidt rank to be controlled unitaries. Following the arguments presented in the previous sections, we can give a sufficient condition that any given unitary is locally equivalent to a controlled unitary. Write unitary $\UC$ as a sum of product operators,
\begin{align}\label{eqn26}
\UC=\sum_{j=1}^{N}M_j^{(1)}\otimes M_j^{(2)}\otimes\ldots\otimes M_j^{(P)},
\end{align}
so that,
\begin{align}\label{eqn27}
I=\UC^\dag\UC=\sum_{i,j=1}^{N}M_i^{(1)\dag}M_j^{(1)}\otimes M_i^{(2)\dag}M_j^{(2)}\otimes\ldots\otimes M_i^{(P)\dag}M_j^{(P)}.
\end{align}
If for any given $\alpha$, operators $\{M_i^{(\alpha)\dag} M_j^{(\alpha)}\}$ span a two-dimensional space, Lemma \ref{lem2} tells us that $\{M_i^{(\alpha)}\}_{j=1}^{N}$ can all be simultaneously diagonalized by local unitaries $U^{(\alpha)},V^{(\alpha)}$. By Lemma \ref{lem1}, we then have that each such party can act as a control for this unitary. Therefore,
\begin{thm9}\label{thm9}
Given nonlocal unitary $\UC=\sum M_j^{(1)}\otimes\ldots\otimes M_j^{(P)}$ acting on $P$ parties and having any Schmidt rank, if for any given $\alpha$, operators $\{M_i^{(\alpha)\dag} M_j^{(\alpha)}\}$ span a two-dimensional space, that party can act as a control for $\UC$, and this is true for every such party.
\end{thm9}

\section{Conclusions}\label{conc}
We have shown that every nonlocal multipartite unitary having Schmidt rank equal to $2$ is locally equivalent to a fully controlled unitary with all parties but one acting as a control.  In the bipartite case we get a stronger result: any bipartite unitary of Schmidt rank $2$ is locally equivalent to a controlled unitary where either party can be chosen as the control, and at least one party can control with two terms, which implies that such unitary can be implemented using LOCC and a maximally entangled state on two qubits. We also provided a sufficient condition for when a nonlocal unitary on any number of parties is locally equivalent to a controlled unitary, and this condition allows one to identify which, if any, parties can act as controls.

As mentioned in the Introduction, our main results can be applied to the gate decomposition (quantum circuit design) of unitaries that act on several systems in the same laboratory. For example, the bipartite result would imply that any unitary acting on systems $A$ and $B$ with Schmidt rank 2 can be expressed as $\UC=(U^{(A)}_1 \otimes U^{(B)}_1) Q (U^{(A)}_2 \otimes U^{(B)}_2)$, where $U^{(A)}_j$ and $U^{(B)}_j$ are local unitaries on $A$ or $B$, and $Q$ is a controlled unitary.

A natural extension of the work presented in this paper on unitaries of Schmidt rank-2 would be to characterize higher Schmidt rank nonlocal unitaries in terms of product operator expansions, beginning with multipartite unitaries of small Schmidt rank.  Such studies may help us better understand the entanglement cost of implementing nonlocal unitaries using LOCC protocols.  

\acknowledgments SC was supported in part by the National Science Foundation through Grant No. PHY-1205931.  L. Yu was supported by the National Research Foundation and Ministry of Education in Singapore.

\appendix*
\section{Proof of Lemma \ref{lem2}}
Since the identity operator lies in the two-dimensional span of operators $\{R_i^\dag R_j\}$, there exist indices $K,L$ such that for all $i,j$, $R_i^\dag R_j$ lies in the span of $\{I,R_K^\dag R_L\}$. In particular, we can write
\begin{align}\label{eqnA1}
R_i^\dag R_L=\mu_iI+\nu_iR_K^\dag R_L~\forall{i}.
\end{align}
Suppose we fix $K$ and seek $L$ such that $\exists{i}$ for which $\mu_i\ne0$. If this is impossible, then 
\begin{align}\label{eqnA4}
R_i^\dag R_j=\hat\nu_{ij}R_K^\dag R_j~ \forall{i,j}.
\end{align}
However, this means
\begin{align}\label{eqnA3}
I=\sum_{i,j}\alpha_{ij}R_i^\dag R_j=R_K^\dag\sum_{i,j}\alpha_{ij}\hat\nu_{ij}R_j,
\end{align}
which implies that $R_K$ is full rank, and then with $j=K$ in \eqref{eqnA4}, we have that every $R_i$ is proportional to $R_K$. This is impossible given that $\{R_i^\dag R_j\}$ spans a two-dimensional space, so for each $K$ there exists a choice of $L$ such that $\mu_i\ne0$ for at least one $i$ in \eqref{eqnA1}, and we immediately see that $R_L$ is full rank. Hence $R_L^{-1}$ exists, and from \eqref{eqnA1} we have
\begin{align}\label{eqnA2}
R_i^\dag=\mu_i R_L^{-1}+\nu_iR_K^\dag~\forall{i}.
\end{align}
Setting $i=L$ and choosing unitaries $U,V$ such that $UR_LV^\dag$ is diagonal, we see that $UR_KV^\dag$ is also diagonal, unless $\nu_L=0$. If $\nu_L\ne0$, then by \eqref{eqnA2}, $UR_iV^\dag$ is diagonal for all $i$, and we are finished.

Suppose now that there is no choice of $K,L$ such that $\nu_L\ne0$. Then $\exists L$ such that $\nu_L=0$, and this implies that $R_L=\sqrt{\mu_L}W_L$ with $W_L$ unitary and $\mu_L\ne0$% and $\mu_L\in\RC$
. Then, from \eqref{eqnA1}
\begin{align}\label{eqnA5}
\sqrt{\mu_L}R_i^\dag W_L=\mu_iI+\nu_i\sqrt{\mu_L}R_K^\dag W_L~\forall{i},
\end{align}
and if $R_K$ is proportional to a unitary, we can choose unitary $V$ such that $VR_K^\dag W_LV^\dag$ is diagonal, in which case $VR_i^\dag W_LV^\dag$ is diagonal $\forall{i}$ and choosing $U=VW_L^\dag$, we are done.

Finally, consider the case that there is no choice of $K,L$ such that $\nu_L\ne0$ and $R_K$ is not proportional to a unitary. Then, $R_K^\dag R_K$ is not proportional to the identity operator, and we can write
\begin{align}\label{eqnA6}
R_i^\dag R_j=\mu_{ij}I+\nu_{ij}R_K^\dag R_K~\forall{i,j}.
\end{align}
We still have $L$ such that $R_L=\sqrt{\mu_L}W_L$, as above. Then,
\begin{align}\label{eqnA6}
\sqrt{\mu_L}R_i^\dag W_L=\mu_{iL}I+\nu_{iL}R_K^\dag R_K~\forall{i}.
\end{align}
Choosing unitary $V$ so that $VR_K^\dag R_KV^\dag$ is diagonal, we then have that $VW_L^\dag R_iV^\dag$ is diagonal $\forall{i}$. Choosing $U=VW_L^\dag$ completes the proof.\hspace{\stretch{1}}$\blacksquare$

%\bibliography{QInforefs}
%\bibliographystyle{prsty}

\end{document}